\documentclass{article}

\usepackage{amssymb}
\usepackage{amsmath}    % allows to have nonnumbered equations and declare new math operators

\usepackage{natbib}

\usepackage{graphicx}
\usepackage[singlelinecheck=false]{caption}

\usepackage{multirow}

\usepackage[usenames, dvipsnames]{color}    % more text coloring options
\usepackage[normalem]{ulem}

\usepackage{hyperref}
\hypersetup{colorlinks,%
citecolor=black,%
filecolor=black,%
linkcolor=black,%
urlcolor=black,%
citebordercolor={0 1 0},
linkbordercolor={0 1 0},
}

\usepackage{cleveref}   % allows references with automatically assigned type (section, equation, figure, etc.)

\usepackage{supertabular}   % long tables

\DeclareMathOperator*{\argmin}{arg\,min}

\usepackage{authblk}    % author-affiliation 

\date{}
\begin{document}

\title{Automatic Identification of Twin Zygosity in Resting-State Functional MRI}
\author[1]{Andrey Gritsenko} %\thanks{gritsenko@wisc.edu}}
\author[2]{Martin A. Lindquist}
\author[1]{Gregory R. Kirk}
\author[1]{Moo K. Chung} %\thanks{mkchung@wisc.edu}}
\affil[1]{\small %Department of Biostatistics and Medical Informatics, 
University of Wisconsin, Madison, WI, USA}
\affil[2]{%Department of Biostatistics, 
Johns Hopkins University, Baltimore, MD, USA}
%\affil[3]{Waisman Laboratory for Brain Imaging and Behavior, University of Wisconsin, Madison, WI, USA}

\maketitle

\begin{abstract}
A key strength of twin studies arises from the fact that there are two types of twins, monozygotic and dizygotic, that share differing amounts of genetic information. 
Accurate differentiation of  twin types allows efficient inference on genetic influences in a population. 
However, identification of zygosity is often prone to errors without genotying.
In this study, we propose a novel pairwise feature representation to classify the zygosity of twin pairs of resting state functional magnetic resonance images (rs-fMRI). For this,  
we project an fMRI signal to a set of  basis functions and use the projection coefficients as the compact and discriminative feature representation of noisy fMRI. We encode the relationship between twins as the correlation between the new feature representations across brain regions.
We employ hill climbing variable selection to identify brain regions that are the most genetically affected. The proposed framework was applied to 208 twin pairs 
and achieved 94.19\% classification accuracy in automatically identifying the zygosity of paired images.


\end{abstract}

\section{Introduction}
\label{sec:Intro}

The extent by which genetic factors shape brain function is still largely unknown. Twin brain imaging studies provide a valuable information for quantifying such extend \textit{in-vivo}. 
The power of twin studies arises from the fact that there are only two types of twins, identical (or \textit{monozygotic}, MZ) and fraternal (or \textit{dizygotic}, DZ), that share differing amounts of genetic information. In average, MZ twins are expected to share 100\% of genes, and DZ twins are expected to share only 50\% of genes~\citep{Twins}. By comparing the similarity between MZ and DZ twins, we can quantify the genetic influence in a population.

Unfortunately, zygosity identification of twins is prone to errors even for the obstetricians delivering babies. As many as twenty percent of all twin births are misidentified according to~\citet{mtfs}. 
Recently, a dataset containing high-quality brain images of twins has become available through the Human Connectome Project (\url{http://www.humanconnectome.org}, HCP).  
In HCP, 35 pairs originally self-reported as DZ twins were later confirmed to be MZ twins after genotyping. Out of 243 twin pairs, this produces the error rate of 14\%. Such a high misclassification rate most likely contributed significantly to mislabeling in many past twin studies without genotyping. In this paper, we explore the feasibility of developing a reliable pipeline for automatic zygosity classification without genotyping.

The significance of genetic contribution in twin studies has been reliably shown for a wide range of functional brain imaging studies ~\citep{TwinStudy2015,TwinStudy2016}, including  heritability of neural activation during simple visuomotor task~\citep{PARK20121132}, calculation~\citep{PINEL2013306},  oral reading recognition and picture vocabulary comprehension~\citep{Feremi2017}, estimating genetic contribution to brain activation in neural networks supporting working memory tasks~\citep{KARLSGODT2007191,BLOKLAND200870,Blokland10882,Koten1737}. Several studies involving resting state functional MRI have also discovered significant genetic contributions to functional network connectivity architecture of the human brain~\citep{TwinStudy2010,Fornito3261,VANDENHEUVEL201319,Gao11288,Yang2016}. These studies mainly utilized the concept of functional connectivity to infer the heritability of brain regions using the Pearson's correlation coefficient between blood-oxygen-level dependent (BOLD) contrasts. Inference on a region's heritability is then typically performed using the ACE model, which factors all differences in population to three components: additive genetic effects, common and unique environments~\citep{ACEmodel,HI}. 
In this study, we design a new framework to determine heritability of brain regions in a model-free setting.

The idea of using functional MRI data for individual identification and prediction has been successfully applied in several recent studies. For example, \citet{miranda2014connectotyping} described a model-based approach capable of identifying a functional fingerprint, 'connectotype', in individual participants through predicting resting state brain activity of each ROI as the weighted sum of all other ROIs. 
\citet{finn2015fingerprinting} investigated this further and showed that individual connectotype of brain activity is preserved not only across scan sessions but also between task and rest conditions. \citet{smith2015positive} used Canonical Correlation Analysis (CCA) to demonstrate a strong covariation link between brain connectivity and demographics, psychometrics and behavioral subject measures.
However, these approaches are not directly applicable in their original form to our specific problem of zygosity classification. In these papers, the main aim is to accurately distinguish each subject from the rest of population, while our goal is a more general classification scheme of classifying paired images into two categories. 

In the past, machine learning methods such as artificial neural networks (ANN), support vector machines, $k$-nearest neighbor, Gaussian na{\"\i}ve Bayes and fuzzy classifiers~\citep{LACONTE2005317,singh2007detection,PEREIRA2009S199,Peltier5332805,Vergun2013,Wang2014,honorio2015classification,wang2017depression} have been used to analyze fMRI data. However, none of them have been used for classifying paired twin images. 
It is unclear how to apply existing classifiers to twin rs-fMRI. For the first time, we propose a unified classification framework that automatically determines the zygosity of a twin-pair using resting-state fMRI and identifies specific brain regions with significant genetic influence. The framework combines the algebraic representation of fMRI time series at the voxel-level with the sparse version of a multi-layer artificial neural network.

For our non-trivial classification problem setting, where each classification identity is represented by a pair of twin images, a special type of neural network is required, specifically designed to process paired data through a comparative analysis. This type of neural networks is related to siamese neural network (SNN)~\citep{Siamese1,zagoruyko2015learning} and has recently gained interest in the medical imaging community~\citep{kouw2017mr,Ktena2017,WANG201812,KTENA2018431}.
One of the main features of SNN is the ability to learn the suitable embedding space of the original data, that is usually composed of highly abstract and semantic information. The drawback of SNN is that  it is difficult to obtain a meaningful biological interpretation of this abstract embedding.  
Instead of learning paired representation through a blackbox approach like SNN, we propose to introduce a parametric algebraic framework that provides an easy-to-comprehend twin representation. In our framework,  a new compact feature representation for the fMRI signal at each voxel is obtained using the cosine series representation (CSR). The correlation between twins at voxel level computed using CSR features is feed into in a two-layer ANN. 
Currently, the majority of brain image analysis studies compute correlation directly from resting state fMRI signal~\citep{rsfMRI1,PELTIER200510,ROGERS20071347,Liang2012Correlation,finn2015fingerprinting,brainsync}. 
In contrast, CSR compactly represent fMRI components into frequency components, which provides a superior performance with respect to the classification accuracy. Our framework can further incorporate a variable selection procedure and able to localize the brain regions that are most heritable. 

\begin{figure*}[t]
\centering
    \includegraphics[width=\textwidth]{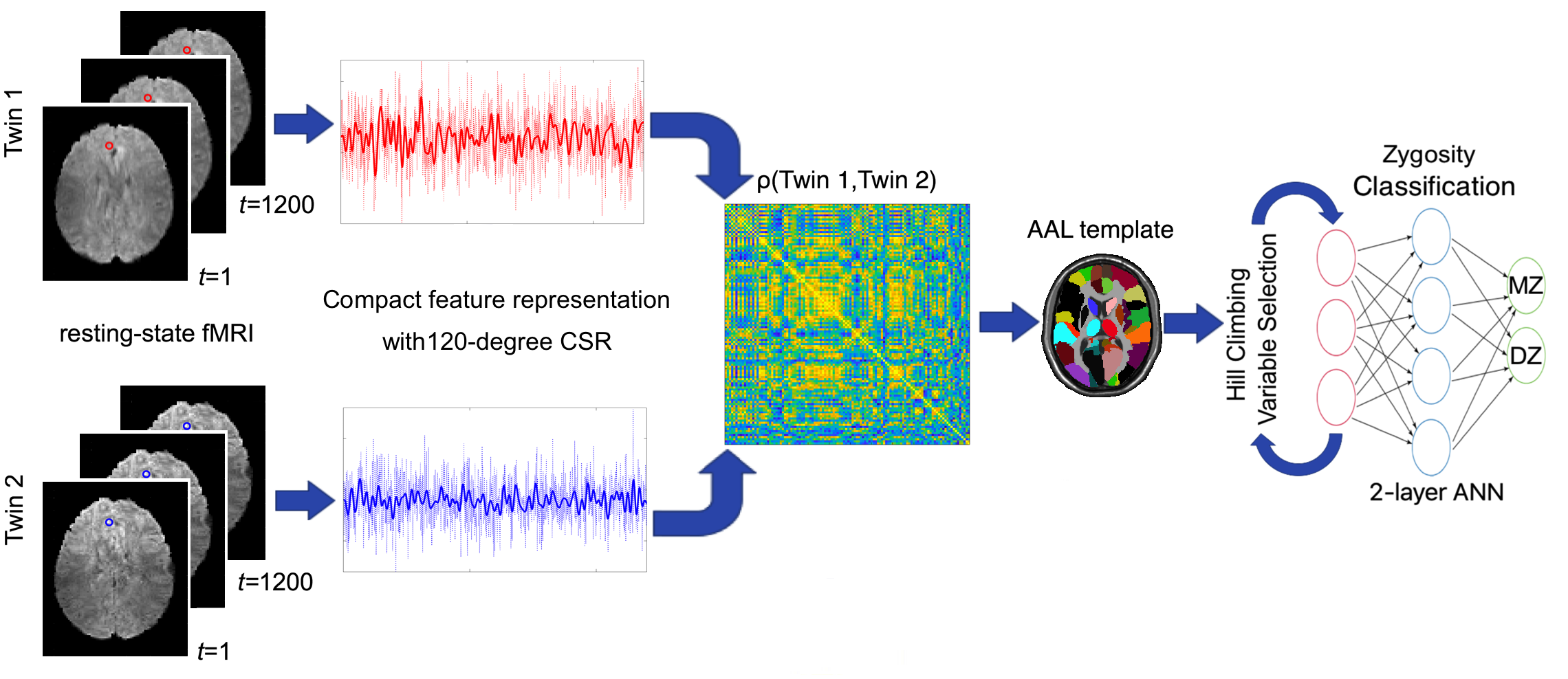}
    \caption{Pipeline of the proposed framework. The 120-degree cosine series representations are displayed as the solid lines.  Twin correlation between CSR coefficients are computed at each voxel. We use AAL brain atlas to parcellate the voxel-level correlations into 116 region-level correlations.
The resulting vectors of 116 twin correlations are feed into a two-layer feedforward neural network to automatically determine zygosity of twin pair.
    }
    \label{fig:Framework}
\end{figure*}

The main contributions of the paper are: 
(1) we introduce a new feature representation in characterizing twin-wise relationships in the whole brain that substantially improves classification performance;  (2) for the first time, we develop a highly reliable zygosity classification scheme without genotyping with $94.19\%$ accuracy;
(3) we propose a principled way of determining the most genetically heritable regions in the brain through the classification scheme.

The proposed framework is applied to HCP, one of the largest publicly available twin datasets, that contains rs-fMRI data of 131 MZ and 77 DZ twin pairs.

\section{Material and Methods}
\label{sec:Methods}

\subsection{Dataset}
\label{sec:Data}

We use resting-state fMRI scans collected as part of the Washington University-Minnesota Consortium Human Connectome Project (HCP) ~\citep{VANESSEN20121299,HCP,VANESSEN201362}.
All participants gave informed consent. The HCP dataset provides information about both self-reported and genotyping-verified zygosity of each twin pair. 
Subjects' genotyping data has been derived from blood or saliva based genytyping~\citep{HCP1200}. 
As our interest lies in identifying the zygosity of twin pairs, we only investigated fMRI scans of genetically confirmed 149 monozygotic (MZ) and 94 same-sex dizygotic (DZ) twins. The zygosity status has not been confirmed for 19 MZ and 9  DZ twin pairs via genotyping. Out of genetically-verified 243 pairs, we excluded 35 twins with missing functional MRI data resulting in the final dataset consisting of 131 MZ (age $29.3\pm3.3$, 56M/75F) and 77 DZ twin pairs (age $29.1\pm3.5$, 30M/47F). Demographic details of twin pairs are summarized in~\Cref{tab:Data}.

\begin{table}[t]
\caption{Demographic data of twin groups}
\centering
\renewcommand{\tabcolsep}{6pt}
\begin{tabular*}{\linewidth}{@{\extracolsep{\fill}\hspace{6pt}} c  c  c  c}
\hline
\rule{0pt}{12pt}Twin group & Sample size & Sex (M/F) & Age \\
\hline
MZ & 131 & 56/75 & $29.3\pm3.3$ \rule{0pt}{12pt} \\ 
DZ & 77 & 30/47 & $29.1\pm3.5$ \\ 
\hline
\end{tabular*}
\caption*{MZ: monozygotic twins, DZ: dizygotic twins. Age is displayed in years as Mean$\pm$SD.}
\label{tab:Data}
\end{table}

All subjects were scanned on a customized Siemens 3T Connectome Skyra scanner housed at Washington University in St. Louis, using a standard 32-channel Siemens receive head coil and a customized SC72 gradient insert and a customized body transmitter coil with 56 cm bore size. 
Resting-state functional MRI were collected over 14 minutes and 33 seconds using a gradient-echo-planar imaging (EPI) sequence with multiband factor 8, time repetition (TR) 720 ms, time echo (TE) 33.1 ms, flip angle $52^\circ$, $104\times90$ (RO$\times$PE) matrix size, 72 slices, 2 mm isotropic voxels, and 1200 time points. During each scanning participants were at rest with eyes open with relaxed fixation on a projected bright cross-hair on a dark background and presented in a darkened room~\citep{HCP1200}.

We used fMRI scans that undergone spatial~\citep{GLASSER2013105} and temporal~\citep{SMITH2013144} preprocessing correcting including
: removal of spatial distortions, realigning volumes to compensate for subject motion, 
registering the fMRI data to the structural MNI template,
minimal highpass temporal frequency filtering,
independent component analysis (ICA)-based artifact removal. The resulting volumetric data contains resting-state functional time series with $91\times109\times91=902629$ 2-mm isotropic voxels at 1200 imaging volumes.

\subsection{Cosine Series Representation}
\label{sec:CSR}

Given the fMRI time series $\zeta(\nu,t)$ at the voxel $\nu$ and time $t$, we scale it to the unit interval $[0,1]$, ans then subtract its mean over time $\int_0^1\zeta(\nu,t)\;dt$. The scaled and translated time series is subsequently represented as 
$$\zeta(\nu,t) = \sum_{l=0}^k c_{l,\nu}\psi_l(t), \; t \in [0,1],$$
where $\psi_0(t) =1, \psi_l(t) = \sqrt{2} \cos ( l \pi t)$ are orthonormal cosine basis functions satisfying
\begin{equation*}
    \big\langle\psi_{l},\psi_{m} \big\rangle = \int_0^1\psi_{l}(t)\psi_{m}(t)dt=\delta_{l m}
\end{equation*}
where $\delta_{lm}$ is the Kronecker's delta. The coefficients  $c_{l,\nu}$ are  estimated in the least squares fashion, i.e.,
\begin{equation*}
    c_{l,\nu} = \argmin_{c_{l,\nu} \in\mathbb{R}}\Big\|\sum_{l=0}^k c_{l,\nu}\psi_l(t)-\zeta(\nu,t)\Big\|^2.
\end{equation*}

For our study, the expansion degree $k=119$ is used such that fMRI is compressed into 10\% of its original data size.
The choice of $k=119$ expansion increases the signal-to-noise ratio (SNR) as measured by the ratio of variabilities by 73\% in average over all voxels in 416 subjects, i.e., SNR = 1.73. 
The resulting coefficient vector ${\bf c}_\nu = (c_{0,\nu}, c_{1,\nu}, \cdots, c_{k,\nu})$ is then used to represent the fMRI in voxel $v$. 
The CSR for multiple time series can be computed by solving for $p=1200$ time points
\begin{align*}
    \underbrace{\begin{pmatrix}
        \zeta(\nu_1,t_1) & \cdots & \zeta(\nu_n,t_1) \\
        % \zeta_1(t_2) & \cdots & \zeta_n(t_2) \\
        \vdots & \zeta(\nu_i,t_j) & \vdots \\
        \zeta(\nu_1,t_p) & \cdots & \zeta(\nu_n,t_p) 
    \end{pmatrix}}_{  \mathrm{Z}_{p\times n}} = 
    \underbrace{\begin{pmatrix}
        \psi_0(t_1) & \cdots & \psi_k(t_1) \\
        % \psi_0(t_2) & \cdots & \psi_k(t_2) \\
        \vdots & \psi_l(t_j) & \vdots \\
        \psi_0(t_p) & \cdots & \psi_k(t_p) 
    \end{pmatrix}}_ {\Psi_{p\times k}}
    \underbrace{
    \begin{pmatrix}
        c_{0,\nu_1} & \cdots & c_{0,\nu_n} \\
        % c_{11} & \cdots & c_{1n} \\
        \vdots & c_{l,\nu_i} & \vdots \\
        c_{k,\nu_1} & \cdots & c_{k,\nu_n} 
    \end{pmatrix} \nonumber}_{\mathrm{C}_{k\times n}}.
\end{align*}
The least squares estimate of $C$ is given by
\begin{equation*}
    \widehat{\mathrm{C}} =  (\Psi^T\Psi)^{-1}\Psi^T\mathrm{Z}.
\label{eq:CSR}
\end{equation*}

\subsection{Correlations between CSR}
\label{sec:CSRcorr}

Consider the CSR of fMRI obtained from the first and second subjects in a twin pair at the same voxel $\nu$ (\Cref{fig:Framework})
$$\zeta(\nu,t) = \sum_{l=0}^k \zeta_{l} \psi_l(t),\quad \eta(\nu,t) = \sum_{l=0}^k \eta_{j}\psi_l(t),$$
where
$$\zeta_{l}=\big\langle\zeta,\psi_l\big\rangle,\quad \eta_{l}=\big\langle\eta,\psi_l\big\rangle$$
are CSR coefficients. 
Since fMRI was normalized with respect to the mean over time, the variance of $\zeta(\nu,t)$ and $\eta(\nu,t)$ is given by
$$\sigma^2\!\:\zeta = \int_0^1 \zeta^2(\nu,t)\!\:dt,\quad \sigma^2\!\:\eta = \int_0^1 \eta^2(\nu,t)\!\:dt.$$
The correlation $\rho(\zeta,\eta)$ between $\zeta(\nu,t)$ and $\eta(\nu,t)$ is then given by
\begin{equation*}
    \rho(\zeta,\eta) 
    = \frac{\int_0^1\zeta(\nu,t)\!\:\eta(\nu,t)\!\:dt}{\sigma\zeta(\nu,t)\cdot\sigma\eta(\nu,t)} 
    = \frac{\sum_{l_1=0}^k \sum_{l_2=0}^k   \zeta_{l_1}\eta_{l_2}\big\langle\psi_{l_1},\psi_{l_2} \big\rangle}{\big[\sum_{l=0}^k\zeta_{l}^2\big]^\frac{1}{2}\cdot\big[\sum_{l=0}^k \eta_{l}^2\big]^\frac{1}{2}}
    = \boldsymbol{\zeta}^T\boldsymbol{\eta}, 
\end{equation*}
where 
$$\boldsymbol{\zeta}  = \frac{(\zeta_{1,\nu}, \cdots, \zeta_{k,\nu})^T}{\big[\sum_{l=0}^k\zeta_{l,\nu}^2\big]^{1/2}},\quad \boldsymbol{\eta} = \frac{(\eta_{1,\nu}, \cdots, \eta_{k,\nu})^T}{\big[\sum_{l=0}^k\eta_{l,\nu}^2\big]^{1/2}}$$
are vectors of cosine series representation coefficients.

Note $\rho$ is the correlation of  low frequency components of time series in the frequency domain. In task-related twin fMRI studies, correlation between twins can be computed in a straightforward manner in the temporal domain because the timing of neuronal activation is comparable across subjects due to the exterior task. However, in the resting-state fMRI there is no external anchor that will lock brain activation of twin subjects across time. 
Thus, we utilize cosine series representation of fMRI signals to compute correlation between twin subjects in the  frequency domain. Similar approaches were used in \citep{curtis.2005,ombao.2008}, where frequency components of signals are correlated using coherence.

\subsection{Twin Classification with Artificial Neural Network}
\label{sec:TwinANN}

Given paired twin images represented as a vector of region-level correlations between CSR coefficients of the original twin fMRI, we use ANN to identify if they belong to a pair of MZ or DZ twins. 
Since we classify the relationship between paired images, this is not a traditional binary classification problem often performed in brain imaging.
We use a two-layer feed-forward ANN, with 200 sigmoid hidden and 1 softmax output neurons to classify the vectors of average correlations~\citep{SCGD}.

We further employ AAL parcellation to compute pairwise twin correlation at region level by
averaging across voxels in each parcellation. Since we cannot average correlations by taking the arithmetic mean without biasing, we transform correlations using the Fisher $z$-transform first:
\begin{equation*}
    z = F(\rho) = \frac{1}{2}\ln{\frac{1+\rho}{1-\rho}}.
\label{eq:Fisher}
\end{equation*}
Then, Fisher transformed correlations are averaged and back projected via the inverse transform~\citep{Fisher2}.
The result is the vector of 116 twin correlations that are feed into the neural network.

Due to possible dependencies between AAL regions, some AAL regions contribute differently to the classification accuracy. We introduce an $\ell_1$-regularization term to the loss function:
\begin{equation}
    \mathcal{L}(w) = -\frac{1}{N}\sum_{i=1}^N[t_i\log y_i + (1-t_i)\log(1-y_i)] + \lambda\sum_{\kappa=1}^M\|w_\kappa\|_1,
\end{equation}
where $t_i$ is the class label for the $i$-th twin pair ('1' for MZ twin pair, '0' for DZ twin pair), $y_i$ is the predicted class label for the $i$-th twin pair, $N$ is the number of twin pairs in the training set, $w_\kappa$ is a vector of weights corresponding to a $\kappa$-th AAL region, and $M=116$ is the number of regions of interest. 
Adding regularization to the ANN prevent model overfitting. Further, $\ell_1$-regularization increases the model's sparsity by penalizing large weights between neurons, thus implicitly removing unimportant features from the model. In brain imaging, regularization is mainly used for segmentation tasks and feature selection~\citep{Liu2018,SANROMA2018143,GUO2017197,WANG201561}. However, here we use it to remove brain regions, whose contribution to zygosity identification can be neglected.

Solving a binary classification problem with artificial neural nets is equivalent to solving a regression problem with subsequent application of the thresholding rule: 
\begin{equation}
    y_i=\begin{cases}
        1\quad\text{if}\;\;o_i\geq\theta, \\
        0\quad\text{if}\;\;o_i<\theta, 
    \end{cases}
\end{equation}
where $o_i$ is numerical output of the neural network for the $i$-th pair of twins, and $\theta$ is the discrimination threshold.

Given a binary classification problem, we take the fMRI data from pairs of MZ twins belonging to the positive class ($t_i=1$) and the fMRI data from pairs of DZ twins belonging to the negative class ($t_i=0$).
We use the holdout method to split dataset randomly into training (70\% of the data), validation (15\%) and test (15\%) subsets. 
The validation dataset is used to avoid overtraining of the model, and to fine tune hyperparameters of the neural network, e.g., the number of hidden neurons and discrimination threshold $\theta$.
Splitting data into training, validation and test subsets 
in the proportion of 70:15:15 is considered as the gold standard~\citep{Datasplit}. 
We trained 1000 independently initialized models without preserved class ratio and average performance across them. We employ classification accuracy, false-positive rate (FPR) and false-negative rate (FNR) to measure the performance of the proposed framework.

\subsection{Boosting Accuracy with Variable Selection}
\label{sec:Boosting}

To infer the contribution of different regions to classification accuracy of the model, we further implemented a hill climbing variable selection procedure considering each AAL parcellation as a variable~\citep{HillClimbing}. 
Hill climbing is an iterative optimization technique that attempts to find a better solution by incrementally changing a single element of the solution. If the change produces a better solution, an incremental change is made to the new solution, repeating until no further improvements can be found.
With the application to variable selection, we implement hill climbing procedure as follows.

We start with the empty variable space and a pool of candidate variables. 
At each iteration, we test candidate variables by picking one variable at a time, adding it to the model and estimating the performance of the model. When all candidate variables are tested, the variable that provides the best performance, with respect to classification accuracy, false positive rate (FPR) and false negative rate (FNR), is removed from the pool of candidates and added to the variable space of the model. Variables are ranked with respect to classification accuracy. 
We continue the process iteratively until all variables are added to the model. 
We rank AAL regions with respect to the classification accuracy they provide to the model based on iterations, when the corresponding variables are added to the variable space. The higher the region is ranked by hill climbing, the more the region is affected by the genetic effect.

\section{Simulation Study}
\label{sec:Simulation}

\begin{table}[t]
\caption{Simulation results}
\centering
\renewcommand{\tabcolsep}{15pt}
\begin{tabular*}{\linewidth}{c  c  c  c}
\hline
\multicolumn{1}{c}{\rule{0pt}{12pt}Study} & Accuracy (\%) & FPR (\%) & FNR (\%) \\
\hline
 \emph{Study 1} & 48.63($\pm$12.37) & 50.64($\pm$24.24) & 49.98($\pm$24.04) \\
 \emph{Study 2} & 79.79($\pm$10.66) & 24.65($\pm$14.66) & 13.51($\pm$14.43) \\ 
 \emph{Study 3a} & 81.88($\pm$10.43) & 23.02($\pm$16.84) & 12.98($\pm$14.40) \\
 \emph{Study 3b} & 87.65($\pm$8.11) & 16.93($\pm$13.98) & 7.24($\pm$9.78) \\
\hline
\end{tabular*}
\caption*{FPR (false-positive rate), FNR (false-negative rate) and overall classification accuracy are provided with standard deviation. 
Study 1: No twin difference. Study 2: Twin difference.
For Study 3: Differential twin difference, we present classification performance both not employing (Study 3a) and employing (Study 3b) variable selection procedure. 
 }
\label{tab:Simulation}
\end{table}

We validated the proposed twin classification framework using simulation studies with known ground truth. 
We first generated ground truth data that represents the underlying resting-state functional MRI signals from $M=5$ distinct brain regions of interest using degree 5 CSR. The CSR coefficients are
$$\mathbf{c_1} = \mathbf{c_2} = \mathbf{c_3} = \mathbf{c_4} = \mathbf{c_5} = \Big[1, \frac{1}{2}, \frac{1}{3}, \frac{1}{4}, \frac{1}{5} \Big]^T.$$
We generate twin data in $\kappa$-th region using the mixed-effect model:
\begin{equation}
\begin{gathered}
    \zeta_{\kappa,i} = \mathbf{c_\kappa} + \alpha_{\kappa,Twin} + \beta_{\kappa,i},\\
    \eta_{\kappa,i} = \mathbf{c_\kappa} + \alpha_{\kappa,Twin} + \beta_{\kappa,i},
\end{gathered}
\end{equation}
where $\zeta_{\kappa,i}$ is the vector of CSR coefficients of the first twin in the $i$-th pair, $\eta_{\kappa,i}$ is the vector of CSR coefficients of the second twin in the $i$-th pair, $\alpha_{\kappa,Twin}$ is a twin-level noise distributed as $N(0,\sigma_{\kappa,Twin}^2)$ and $\beta_{\kappa, i}$ is an individual-level noise distributed as $N(0,\sigma_{\kappa,Ind}^2)$. 
After the synthetic twin data is generated, we applied the proposed ANN framework: compute correlations between twins, train ANN to classify zygosity of twin pairs, and estimate the extent by which regions of interest are affected by the genetic effects using hill climbing variable selection procedure. 

\begin{figure}[t]
\centering
    \includegraphics[width=\linewidth]{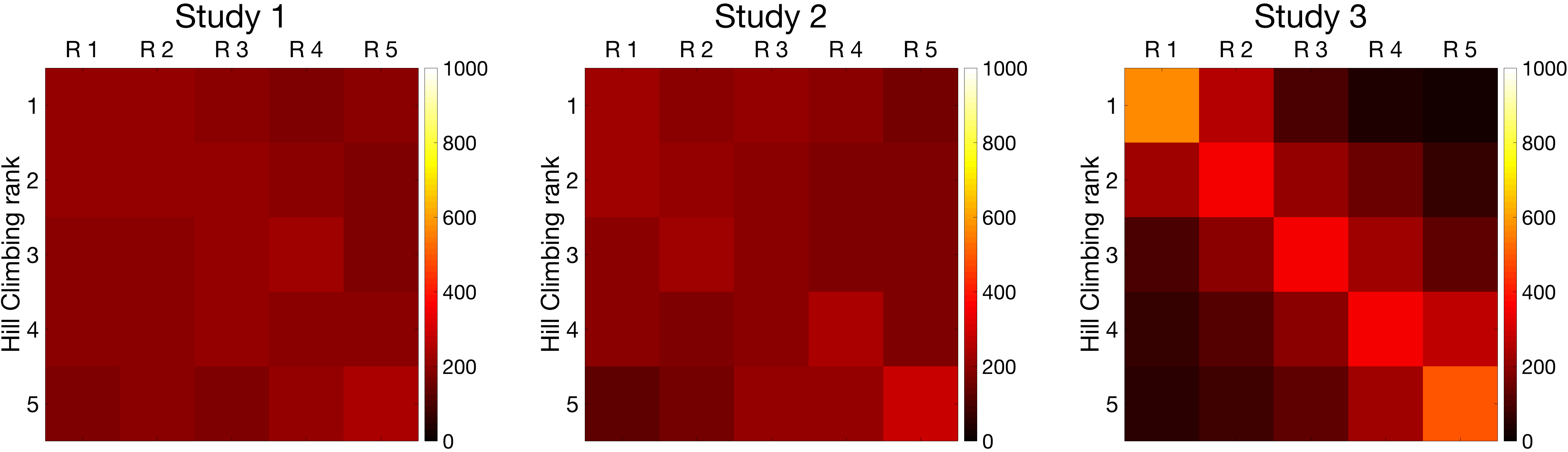}
    \caption{Results of hill climbing variable selection for three simulation studies. 
    For each region of interest $\kappa$, 
    The value at the intersection of $\kappa$-th column and $i$-th row represents how often this regions has been selected at the $i$-th iteration of the hill climbing variable selection procedure.     }
\label{fig:HCranking-Simulation}
\end{figure}

We perform three different simulation studies using the same ground truth coefficients $\mathbf{c_\kappa}$ and individual-level variance $\sigma_{\kappa,Ind}^2 = 0.25^2$, but different twin-level variances, $\sigma_{\kappa,MZ}^2,\,\sigma^2_{\kappa,DZ}$ depending on the zygosity of twin pairs. In all three simulations, we generate 50 MZ pairs and 50 DZ pairs. To obtain stable results we repeated the simulation 1000 times and the average results are reported. The simulation results are processed simultaneously by logistic regression model and our classification framework and averaged results are summarized in \Cref{tab:Simulation} and \Cref{fig:HCranking-Simulation}. 

\paragraph{Study 1: No twin difference} We tested if the method is detecting any false positive when there is no twin difference.  
By  letting the DZ-variability equal the MZ-variability
$$\sigma_{\kappa,DZ}^2 = \sigma_{\kappa,MZ}^2 = \sigma_{\kappa,Ind}^2,$$
we are generating twin data without any DZ- and MZ-twin difference. 
The classification accuracy for the proposed framework is $48.63 \pm12.37\%$ indicating we are not falsely classifying the zygosity.

\paragraph{Study 2: Twin difference} We forced all 5 regions to be equally highly heritable. We achieve this by simulating MZ twin data with smaller twin-level variance compared to DZ twin data:
$$\sigma_{\kappa,MZ}^2 = \sigma_{\kappa,Ind}^2,\,\sigma_{\kappa,DZ}^2 = 2^2 \sigma_{\kappa,Ind}^2,$$
The classification accuracy for the proposed framework is $79.79 \pm10.66\%$ indicating we are classifying zygosity. 

\paragraph{Study 3: Differential twin difference}
We forced 5 regions to have gradually decreasing heritability:
$$\sigma_{\kappa,MZ}^2 = \sigma_{\kappa,Ind}^2,\,\sigma_{\kappa,DZ}^2 = h_\kappa^2 \sigma_{\kappa,Ind}^2,$$
where 
$\{h_\kappa\} = \{3,2.5,2,1.5,1\}$ is a sequence of gradually decreasing numbers that control heritability of each $\kappa$-th region of interest.
The classification accuracy for the proposed framework in Study 3 without hill climbing is $81.88 \pm10.43\%$.
When utilizing hill climbing, the accuracy increased to $87.65 \pm8.11 \%$  indicating we are gaining advantage by employing variable selection in case of unequal heritability across regions. ROI rankings provided by hill climbing correspond to the gradually decreasing heritability of regions (\Cref{fig:HCranking-Simulation}).

\section{Results}
\label{sec:Results}

The proposed ANN framework was applied to the HCP rs-fMRI data from 208 twin pairs. 
The classification performance is reported in~\Cref{tab:Results}. We achieved $54.15(\pm9.24)\%$ accuracy
when original rs-fMRI were used in computing twin correlations. The use of CSR improved the accuracy significantly and achieved $79.93(\pm7.59)\%$ accuracy with $36.98(\pm17.29)\%$ false-positive rate and $9.99(\pm6.79)\%$ false-negative rate.

\begin{table*}[t]
\caption{Performance of the proposed classification pipeline at different stages}
\centering
\renewcommand{\tabcolsep}{3pt}
\begin{tabular*}{\textwidth}{ @{\extracolsep{\fill}\hspace{2pt}} l  c  c  c}
\hline
\multicolumn{1}{c}{\rule{0pt}{12pt}Method} & Accuracy (\%) & FPR (\%) & FNR (\%) \\
\hline
ANN w/o CSR & 54.15($\pm$9.24) & 74.67($\pm$18.74) & 28.36($\pm$17.09) \\ 
ANN w/ CSR & 79.93($\pm$7.59) & 36.98($\pm$17.29) & 9.99($\pm$6.79) \\ 
ANN + Hill climbing & \textbf{94.19($\pm$3.53)} & \textbf{9.54($\pm$6.87)} & \textbf{3.69($\pm$3.47)}\\ 
Logistic regression & 47.99($\pm$3.32) & 61.62($\pm$5.56) & 38.24($\pm$4.14) \rule{0pt}{12pt} \\ 
\hline
\end{tabular*}
\caption*{
FPR: false-positive rate, FNR: false-negative rate.
Overall classification accuracy, false-positive rate and false-negative rate are provided for the test subset in percentages with standard deviations. }
\label{tab:Results}
\end{table*}

\subsection{Detecting the Most Heritable Regions}

\begin{figure*}[t]
\centering
    \includegraphics[width=\textwidth]{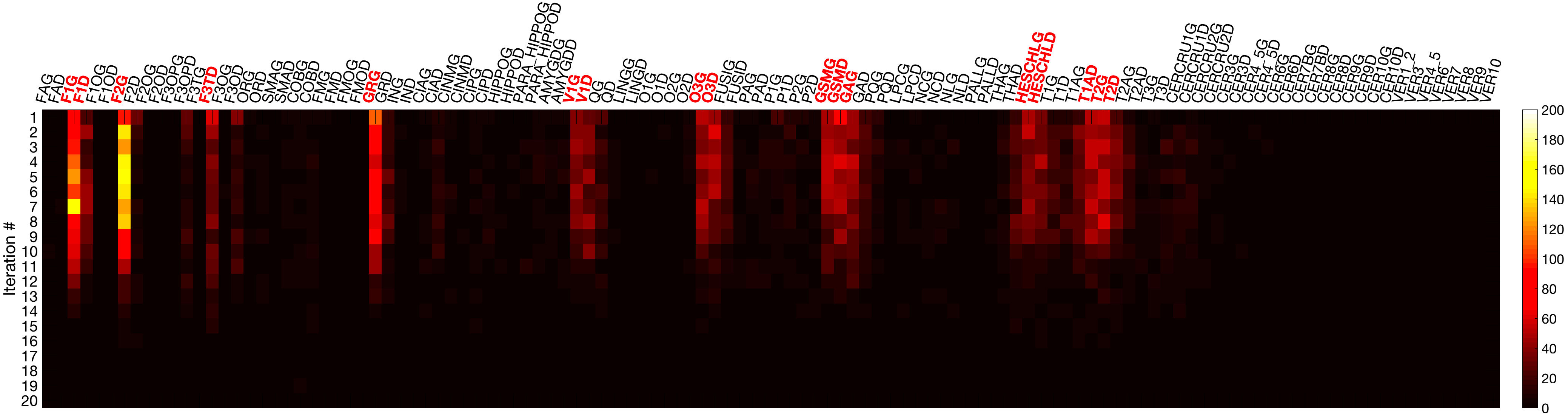}
    \caption{Results of hill climbing variable selection.
    Results are accumulated over 1000 independently initiated models for left-to-right (A) and right-to-left (B) phase encoding. 
    For each region of interest $\kappa$, the value at the intersection of $\kappa$-th column and $i$-th row represents how often this regions has been selected at the $i$-th iteration of hill climbing. 
    As a result, the dimensionality of the variable space is smaller than the original 116-dimensional variable space, and in most cases it consists of 10 AAL regions. 
    Red colored regions represent the variable space of the most important AAL regions (above $85$-th percentile). 
    }
\label{fig:HCranking-All}
\end{figure*}

Hill climbing was used on ANN classification to further identify the most heritable brain regions and improve the performance. 
In hill climbing, each variable represents the average correlation between twins in a single AAL parcellation. 
\Cref{fig:HCranking-All} illustrates the results of the variable selection procedure accumulated across 1000 independently initialized models. \Cref{fig:HCranking-All} displays how often hill climbing selects a certain variable at a given iteration. Based on the results of hill climbing, we infer the extent each ROI contributes to the zygosity classification. 
We estimate the contribution of $\kappa$-th AAL parcellation using the following criterion
\begin{equation*}
    \mathcal{J}(\kappa) = \sum_{i=1}^M  \frac{\gamma(\kappa,i)}{i},
    \label{eq:AALimp}
\end{equation*}
where $\gamma(\kappa,i)$ is the number of times when the brain region $\kappa$ has been added to the variable space at the $i$-th iteration, and $M$ is the total number of the variables ($M=116$ in this study).
Regions that have been selected by hill climbing at least once across all models are marked with bold font, and the most important $85$-th percentile regions are marked in red in \Cref{fig:HCranking-All}.

Employing hill climbing variable selection procedure not only allowed us to estimate the importance of AAL parcellations with respect to the classification accuracy, it also provided us with a tool to find the optimal variable space with the highest possible classification accuracy. The dimensionality of the variable space is smaller than the original 116-dimensional variable space, and in most cases it consists of 10 AAL regions. 
The average performance  of hill climbing was $94.19(\pm3.53)\%$ classification accuracy with $9.54(\pm6.87)\%$ FPR and $3.69(\pm3.47)\%$ FNR.

\begin{figure*}
\centering
    \includegraphics[width=\textwidth]{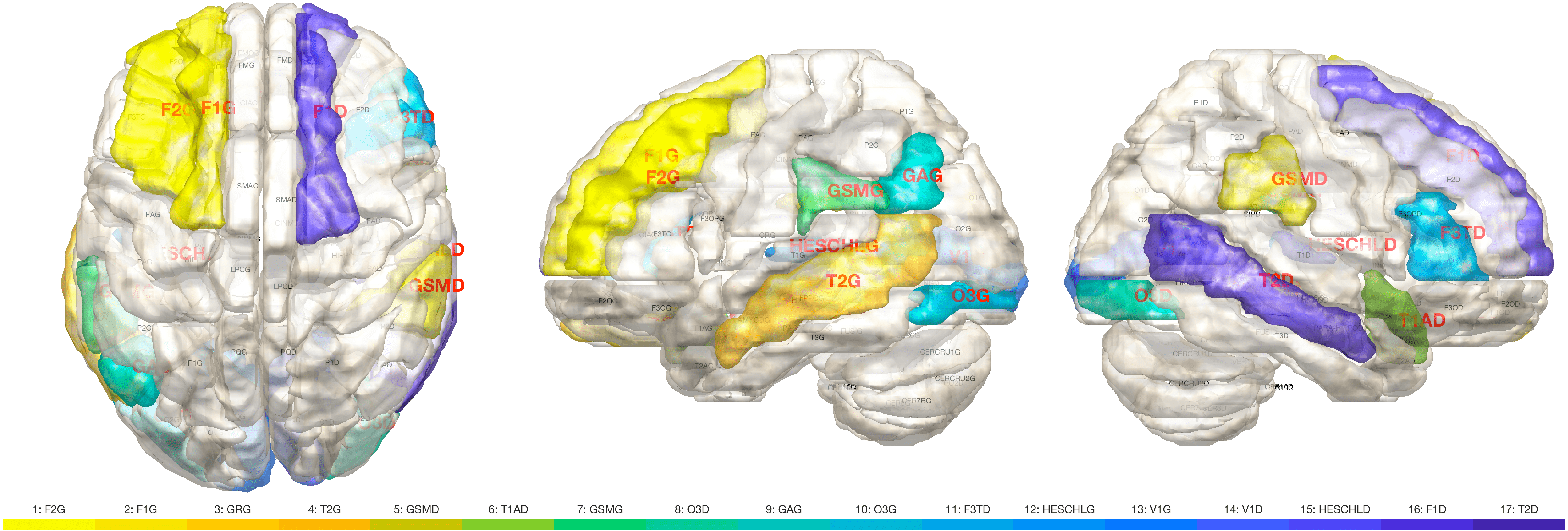}
    \caption{Most heritable brain regions  are ranked according to their importance $\mathcal{J}(\kappa)$ and   $85$-th percentile regions are colored, which corresponds to regions marked with red font in \Cref{fig:HCranking-All}. Refer to~\Cref{tab:FirstHC-LR_RL} for a detailed information on the contribution of each parcellation to the performance of classification model.
   }
\label{fig:TopHI-LR_RL}
\end{figure*}

The following areas are identified as the most important AAL regions with respect to the overall contribution to the classification accuracy (Figure \ref{fig:TopHI-LR_RL}):
Left middle frontal gyrus, lateral part; 
Left superior frontal gyrus, dorsolateral; 
Left gyrus rectus; 
Left middle temporal gyrus; 
Right supramarginal gyrus; 
Right superior temporal pole; 
Left supramarginal gyrus; 
Right inferior occipital; 
Left angular gyrus; 
Left inferior occipital; 
Right area triangularis; 
Left transverse temporal gyri; 
Left calcarine sulcus; 
Right calcarine sulcus; 
Right transverse temporal gyri; 
Right superior frontal gyrus, dorsolateral; 
Right middle temporal gyrus.

\subsection{Comparison Against Logistic Regression}
We compared the performance of the proposed pipeline to an often used standard classifier -- logistic regression.
Given input data $\mathbf{X}=[\mathbf{x}_1,\dots,\mathbf{x}_N]$ and class labels $\mathbf{T}$, linear classifier is a model that maps $\mathbf{X}$ onto $\mathbf{T}$, i.e. $\mathbf{T} = f\big(\mathbf{X}^T\mathbf{w})$.
Here, $\mathbf{w}$ is a vector of model parameters, and $f$ is a function that transforms $\mathbf{X}^T\mathbf{w}$ into desired output values. For logistic regression, $f$ is a sigmoid function. 
In our study, the input data $\mathbf{X}$ is correlation between paired fMRI data, and class labels $\mathbf{T}$ represent zygosity. To train a classifier, we need to find $$w = \underset{\mathbf{w}}{\argmin}\sum_i L(t_i,y_i),$$ where $L(t_i,y_i)$ is a log-loss function that measures the discrepancy between the classifier's prediction value $y_i = f(\mathbf{w}^T\mathbf{x}_i)$ and the true class label $t_i$ for the $i$-th training example. 
Logistic regression algorithm finds maximum likelihood estimation of $\mathbf{w}$ using iteratively reweighted least squares~\citep{Murphy2012}:
\begin{equation*}
    \mathbf{w}^{(k+1)} = \left(\mathbf{X}^T\mathbf{S}_k\mathbf{X}\right)^{-1}\mathbf{X}^T \left(\mathbf{S}_k \mathbf{X} \mathbf{w}^{(k)} + \mathbf{T} - \mathbf{Y}^{(k)}\right),
\end{equation*}
where $\mathbf{Y}^{(k)}=[y^{(k)}_1,\dots,y^{(k)}_N]$ is a vector of classifier outputs,  $$y^{(k)}_i=\frac{1}{1+e^{-\mathbf{w}^T\mathbf{x}_i}}, \quad \mathbf{S}_k=\operatorname{diag}(y^{(k)}_i(1-y^{(k)}_i))$$ is a diagonal weighting matrix. 
For the given HCP dataset, we achieved $47.99(\pm 3.32)\%$ classification accuracy with $61.62(\pm 5.56)\%$ false-positive rate and $38.24(\pm 4.14)\%$ false-negative rate employing logistic regression model. Performance wise, the logistic regression was not perform any better than ANN.

\begin{table*}
\caption{Most frequent AAL regions selected at the first iteration of hill climbing}
\centering
\renewcommand{\tabcolsep}{3pt}
\begin{tabular*}{\textwidth}{ @{\extracolsep{\fill}\hspace{0pt}} l  c  c  c  c}
\hline
\multicolumn{1}{c}{Label} & Frequency & Accuracy (\%) & FPR (\%) & FPR (\%) \\
\hline
% Left gyrus rectus & 
GRG & 83/1000 & 86.36 ($\pm$4.06) & 23.04 ($\pm$10.95) & 8.68 ($\pm$5.16) \rule{0pt}{12pt} \\
% Left middle frontal gyrus, lateral part & 
F2G & 70/1000 & 85.03 ($\pm$4.93) & 22.43 ($\pm$9.73) & 10.66 ($\pm$5.59) \\
% Left superior frontal gyrus, dorsolateral & 
F1G & 52/1000 & 81.63 ($\pm$4.97) & 20.74 ($\pm$8.58) & 16.92 ($\pm$7.22) \\
% Right supramarginal gyrus & 
GSMD & 50/1000 & 85.83 ($\pm$6.76) & 22.39 ($\pm$10.6) & 8.87 ($\pm$6.32) \\
% Right area triangularis & 
F3TD & 45/1000 & 86.48 ($\pm$3.82) & 18.98 ($\pm$9.36) & 9.98 ($\pm$6.39) \\
% Left middle temporal gyrus & 
T2G & 40/1000 & 86.67 ($\pm$5.44) & 19.48 ($\pm$11.79) & 10.22 ($\pm$4.97) \\
% Right superior temporal pole & 
T1AD & 39/1000 & 86.09 ($\pm$6.56) & 20.43 ($\pm$13.14) & 10.61 ($\pm$7.68) \\
% Left inferior occipital & 
O3G & 37/1000 & 87.69 ($\pm$5.56) & 19.5 ($\pm$13.02) & 8.46 ($\pm$5.01) \\
% Left supramarginal gyrus & 
GSMG & 36/1000 & \bf 88.41 ($\pm$4.76) & 20.98 ($\pm$12.27) & 7.22 ($\pm$4.36) \\
% Left transverse temporal gyri & 
HESCHLG & 36/1000 & 85.56 ($\pm$7.1) & 22.17 ($\pm$13.67) & 9.85 ($\pm$5.3) \\ % (left Heschl's gyrus) 
% Right inferior occipital & 
O3D & 34/1000 & 88.32 ($\pm$4.78) & \bf 18.14 ($\pm$11.88) & 8.02 ($\pm$5.13) \\
% Left angular gyrus & 
GAG & 34/1000 & 86.94 ($\pm$4.9) & 20.03 ($\pm$8.23) & 7.71 ($\pm$6.7) \\
% Right orbital part of inferior frontal gyrus & 
F3OD & 30/1000 & 85.42 ($\pm$5.24) & 20.48 ($\pm$11.06) & 11.06 ($\pm$6.19) \\
% Left calcarine sulcus & 
V1G & 28/1000 & 87.32 ($\pm$5.11) & 22.11 ($\pm$11.69) & \bf 5.88 ($\pm$4.75) \\
% Right transverse temporal gyri & 
HESCHLD & 27/1000 & 84.77 ($\pm$7.22) & 21.91 ($\pm$12.99) & 11.24 ($\pm$7.92) \\ % (right Heschl's gyrus) 
% Right middle frontal gyrus, lateral part & 
F2D & 25/1000 & 86.64 ($\pm$4.3) & 20.84 ($\pm$11.6) & 9.46 ($\pm$4.23) \\
\hline
\end{tabular*}
\caption*{FPR: false-positive rate, FNR: false-negative rate.
Overall classification accuracy, false-positive rate and false-negative rate are provided  with the standard deviation. They present performance of the model at the first iteration of hill climbing when the variable space contains only one variable. Bold font highlights the best performing region. 
Region names are given in the first column according to the AAL brain atlas specification (\url{http://neuro.compute.dtu.dk/services/brededatabase/index_roi_tzouriomazoyer.html}), while the second column displays regions' labels as they appear on~\Cref{fig:TopHI-LR_RL}. 
Regions are sorted according to the frequency across 1000 independently initialized models.
}
\label{tab:FirstHC-LR_RL}
\end{table*}

\section{Conclusion and Discussion}
\label{sec:Discussion}

For the first time, we addressed the problem of classifying the zygosity of twin-pairs using resting state functional MRI. This is a more complex problem than the usual classification problem of labeling each image into distinct classes. Here, we are interested in learning if the relationship between pairs of images is associated with the zygosity of twins.

There are two practical advantages for using CSR as a new feature representation of twin fMRI. 
First, employing cosine series representation allows to correlate fMRI signals in the frequency domain. 
Furthermore, representing the original fMRI signals as a linear combination of 120 cosine basis functions serves not only as a dimensionality reduction technique but also as a means of denoising high frequency noise. To emphasize the importance of the new feature representation in the performance of classification, we classified twin fMRI without the cosine series representation (CSR) and obtained $54.15(\pm 9.24)\%$ classification accuracy
(\Cref{tab:Results}, first row).  The use of CSR allowed us to increase the classification accuracy to $79.93(\pm 7.59)\%$, which is an increase of $25\%$.
The proposed framework of ANN with hill climbing achieves the classification accuracy of $94.19 (\pm3.53)\%$ in identifying the zygosity of twins.
Performance of the proposed classification pipeline is much higher when compared to the accuracy achieved on the same dataset by a conventional classification model - logistic regression, $47.99(\pm3.32)\%$.

\paragraph{Regions with significant genetic influence}
We report that the most genetically affected brain regions, as measured by their contribution to the classification accuracy of the zygosity type of twins, are mainly located in the temporo-parietal and frontal brain regions. In order to determine if our findings are consistent with previous studies, we examined previous twin imaging studies. Despite some discrepancy, there is a overlap of our findings with results reported in a number of twin studies of both task-based and resting-state functional MRI~\citep{BLOKLAND200870,Blokland10882,Koten1737,TwinStudy2010,PARK20121132,Gao11288,SINCLAIR2015243,Yang2016}. \citet{Cannon3228} found that genetic influences were isolated primarily to polar, inferior, and dorsolateral prefrontal brain areas, and also in the frontal regions. \citet{MATTHEWS2007223} found that dorsal anterior cingulate cortex activation is significantly influenced by genes. \citet{Olli2008}  demonstrated high heritability of medial and dorsolateral prefrontal cortex. \citet{BLOKLAND200870,Blokland10882} found that the inferior, middle, and superior frontal gyri, left supplementary motor area, precentral and postcentral gyri, middle cingulate cortex, superior medial gyrus, angular gyrus, superior parietal lobule, including precuneus, and superior occipital gyri are genetically affected in twins. \citet{Koten1737} observed significant genetic influences on brain activation in visual cortex, temporo-parietal and frontal areas, and anterior cingulate cortex. \citet{PARK20121132} found neural activity in the left visual cortex and left motor cortex were significantly heritable. \citet{SINCLAIR2015243} found that 47 out of 116 AAL regions are significantly heritable which overlap with most of our regions. Compared to previous twin literature, there is a strong consistency in the identifeid heritable brain regions. However disparities suggest that genetic influences may vary with task paradigms, which rs-fMRI is lacking.

\paragraph{AAL parcellation template}
In the proposed framework, we computed pairwise correlation between twin subjects and then averaged them using the predefined AAL parcellation. 
However, from the review of several structural and functional voxel-based twin studies, it is apparent that genetic effects may carry across anatomical boundaries~\citep{Blokland10882,joshi2011contribution,VANSOELEN20123871}. Therefore, voxel-based approaches may have preference in imaging genetic studies over ROI approaches that average measurements across brain parcels. The drawback of voxel-wise approaches in general is an extensive computational load resulting from the computation of pairwise correlations at voxel level. 
We overcame this drawback by utilizing 120-degree cosine series representation that drastically decreases computational time by compact signal representation. We take the advantage of both methods in increasing the classification accuracy. Although we demonstrated relatively high classification accuracy with a provided solution, there is a need for better parcellation method that balances the trade-off between pure voxel-wise and anatomical template-based approaches. We leave this as a future study.

\paragraph{Variable selection}
To quantify the extent of genetic contribution of AAL parcellations, we used  hill climbing -- a greedy search algorithm that considers regions of interest as independent variables and tests one variable at a time. 
Recent studies of both resting-state and task-related fMRI on the functional connectivity of brain regions has revealed that activity in some regions may have strong coherence~\citep{TwinStudy2015,Yang2016}. 
To get a more accurate inference on the optimal variable space, one may consider applying the concept of "connectivity of regions" and performing a group variable selection, i.e., combine variables in groups and test each group as a single instance~\citep{HillClimbing}. 
Additionally, this will reduce computational load of variable selection procedure, whose computational complexity for brute-force approaches that test all possible combinations of variables is $\mathcal{O}(2^n)$~\citep{Yang2016}. Other type of more complex variable selection methods are left as a future study.

\section*{Acknowledgements}
Data were provided by the Human Connectome Project, WU-Minn Consortium (Principal Investigators: David Van Essen and Kamil Ugurbil; 1U54MH091657), which was funded by the McDonnell Center for Systems Neuroscience at Washington University and the 16 NIH Institutes and Centers that support the NIH Blueprint for Neuroscience Research. This work was supported by NIH grants R01 EB022856 and UL1TR002373. We would like to thank Guorong Wu of University of North Carolina, Chapel Hill and Hernando Ombao of King Abdullah University of Science and Technology for valuable discussions and supports.

\bibliographystyle{agsm} 
\bibliography{biblio}

\end{document}